\documentclass[12pt,a4paper,english]{article}
\usepackage[latin1]{inputenc}
\usepackage{amsmath}
\usepackage{setspace}
\usepackage{amssymb}

\makeatletter
\usepackage{graphics}
\usepackage{color}
\newfont{\mcal}{rsfs10 scaled 1200}
\newfont{\bit}{cmbxti10 scaled 1728}

\usepackage{babel}
\makeatother
\begin{document}
\begin{center}{\LARGE Head-on collision}\end{center}{\LARGE \par}

\begin{center}{\LARGE of }\end{center}{\LARGE \par}

\begin{center}{\LARGE ultrarelativistic charges}\end{center}{\LARGE \par}

\vspace{0.5cm}
\begin{center}{\large Peter C. AICHELBURG$^{\dagger}$\footnote[1]{email:pcaich@ap.univie.ac.at},
Herbert BALASIN$^{\ddag}$\footnote[2]{email:hbalasin@tph.tuwien.ac.at}}\end{center}{\large \par}

\begin{center}and\end{center}

\begin{center}{\large Michael KERBER$^{\dagger}$\footnote[3]{email:a9405544@unet.univie.ac.at}}\end{center}{\large \par}

\begin{center}\textit{\small $^{\dagger}$Institut für Theoretische
Physik, Universität Wien, }\end{center}{\small \par}

\begin{center}\textit{\small Boltzmanngasse 5, 1090 Wien, }\end{center}{\small \par}

\begin{center}\textit{\small AUSTRIA}\end{center}{\small \par}

\vspace{0.2cm}
\begin{center}\textit{\small $^{\ddag}$Institut für Theoretische
Physik, TU-Wien,}\end{center}{\small \par}

\begin{center}\textit{\small Wiedner Hauptstraße 8-10, 1040 Wien,}\end{center}{\small \par}

\begin{center}\textit{\small AUSTRIA}\end{center}{\small \par}
\vspace{0.5cm}

\noindent \begin{center}\textbf{\footnotesize Dedication:} {\footnotesize }\begin{minipage}[t]{0.65\columnwidth}%
\begin{spacing}{0.5}
{\footnotesize It is a pleasure to dedicate this article to Vincent
Moncrief on the occasion of his special birthday. }\end{spacing}
\end{minipage}%
\end{center}{\footnotesize \par}

\begin{abstract}
We consider the head-on collision of two opposite-charged point particles
moving at the speed of light. Starting from the field of a single
charge we derive in a first step the field generated by uniformly
accelerated charge in the limit of infinite acceleration. From this
we then calculate explicitly the burst of radiation emitted from the
head-on collision of two charges and discuss its distributional structure.
The motivation for our investigation comes from the corresponding
gravitational situation where the head-on collision of two ultrarelativistic
particles (black holes) has recently aroused renewed interest. 

\vspace*{1cm}\hspace*{8cm}\begin{minipage}[c]{1.0\columnwidth}%
UWThPh--2002--44\\
TUW~~03--10\end{minipage}%

\end{abstract}

\newpage
\subsection*{Introduction}

Sources moving with high velocity, close to the speed of light, are
of general interest in physics, e.g. \textit{\emph{\large }}for studying
phenomena in high energy particle physics or for understanding radiation
from high speed encounters of black holes. Essential simplification
in the description of such phenomena is achieved by boosting the source
to the velocity of light. Because only massless sources can travel
with the fundamental velocity, this limit requires some care, letting
the rest mass tend to zero while at the same time keeping the energy
constant. Such light-like sources can provide a leading-order approximation
for describing massive sources traveling with ultra-relativistic velocities.
In this paper we study classical solutions of Maxwells equations for
special light-like currents, i.e. charges moving with the velocity
of light in Minkowski space-time. Consider, as the simplest example,
a moving point charge. By boosting the charge to higher and higher
velocities the originally spherical symmetric Coulomb field becomes,
as it is well known, deformed and more and more concentrated in the
plane orthogonal to the charges. Finally, in the limit, this deformation
becomes extreme, the field becomes completely concentrated on the
null-hyperplane which contains the light-like trajectory of the charge.
From the mathematical point of view care must be taken when calculating
the electromagnetic field associated light-like currents by the Green-function
method. The reason being, that the trajectory of light-like sources
lie in the characteristic surfaces of the hyperbolic system. Simply
speaking, the light cone of the point where the field is to be evaluated
may intersect the source at null infinity. These contributions however
are essential for obtaining the correct field.

Our main goal here is to present an exact solution of Maxwell's equation
which is the classical analog of pair annihilation in quantum field
theory: Two opposite charged point sources moving with the speed of
light collide head on and produce a burst of electromagnetic radiation.
After the collision no charge but only radiation is present. Of course
this is an idealised situation but does provide an approximation to
what happen in more realistic situations. The description of colliding
charges may also provide better insight for understanding the radiation
emitted by the collision of two black holes as first obtained by D'Eath
\cite{Death}. More recently, the gravitational interaction of two
colliding massless particles \cite{AS} to produce a black hole was
discussed \cite{Gid,Ven}. Since the sources and fields are distributional
in nature, the electromagnetic setting disentangles the distributional
from the geometrical aspects.

\subsection*{1) The EM-pulse}

Since the situation we are interested in has null (lightlike) character
we adapt our coordinates accordingly, i.e.\[
ds^{2}=-2dudv+d\tilde{x}^{2},\]
where $\tilde{x}$ refers to coordinates in the orthogonal space to
the timelike 2-plane spanned by $l^{a}=\partial_{v}^{a}$ and $n^{a}=\partial_{u}^{a}$
\footnote{Here and in the following small latin sub- and superscripts from the
beginning of the alphabet will denote abstract indices referring to
the tensor type only%
}. 

The current of a point-charge $e$ moving at ultrarelativistic (i.e.
the velocity of light) speed along the direction of $l^{a}$ is given
by\begin{equation}
j=e\delta(u)\delta^{(2)}(\tilde{x})\partial_{v}.\label{eq:ASCURR}\end{equation}
The corresponding electromagnetic field is obtained by solving Maxwell's
equations for the potential $A^{a}$ which conveniently reduce to
a wave equation in the Lorenz-gauge ($\partial_{a}A^{a}=0)$\[
\partial^{2}A^{a}=4\pi j^{a}.\]
Formally the potential may be obtained by employing the Green-function
for the wave operator\[
G(x)=\int\frac{d^{4}k}{(2\pi)^{4}}\frac{-1}{k^{2}}e^{ikx}.\]
Here (and in the following) the Fourier-expressions will be understood
in terms of symbolic calculus, i.e. as representing a solution of
the equation\[
\partial^{2}G(x)=\delta^{(4)}(x).\]
For the Fourier-transform of the current (\ref{eq:ASCURR}) 

\[
j^{a}(k)=e2\pi\delta(k^{n})l^{a}\qquad k^{a}=:k^{l}l^{a}+k^{n}n^{a}+\tilde{k}^{a}\]
this yields\[
A^{a}(x)=4\pi e\int\frac{d^{4}k}{(2\pi)^{4}}e^{ikx}\frac{-1}{k^{2}}2\pi\delta(k^{n})l^{a}.\]
Decomposing $k\cdot x$ and $k^{2}$ with respect to $l^{a}$ and
$n^{a}$\[
k^{2}=-2k^{l}k^{n}+\tilde{k}^{2}\qquad kx=-k^{n}v-k^{l}u+\tilde{k}\tilde{x}\]
this reduces to \[
A^{a}(x)=4\pi el^{a}\int\frac{d^{2}\tilde{k}}{(2\pi)^{2}}\frac{dk^{l}}{2\pi}\frac{-1}{\tilde{k}^{2}}e^{-ik^{l}u+i\tilde{k}\tilde{x}}=4\pi e\delta(u)l^{a}\int\frac{d^{2}\tilde{k}}{(2\pi)^{2}}\frac{-1}{\tilde{k}^{2}}e^{-i\tilde{k}\tilde{x}}.\]
The last expression is nothing but the Fourier-transform of the Green-function
for the two-dimensional Laplacian, i.e.\[
\tilde{\partial}^{2}f(\tilde{x})=\delta^{(2)}(\tilde{x})\qquad\tilde{f}=\frac{1}{2\pi}\log\rho,\,\,\,\,\rho^{2}=\tilde{x}^{2}.\]
Therefore the potential for the EM-pulse becomes\[
A^{a}(x)=2e\delta(u)\log\rho l^{a}\]
from which we readily obtain the field-strength\begin{equation}
F=dA=-2e\delta(u)\frac{1}{\rho}d\rho\wedge du.\label{eq:ASfield}\end{equation}
The charge produces an impulsive electromagnetic field (EM-pulse).
As expected this field is concentrated on the null hypersurface $u=0$
and is singular along the trajectory of the charge $u=\rho=0$. In
deriving $F_{ab}$ we have implicitly chosen boundary conditions.
Actually calculating the field explicitly with retarded or advanced
Green-functions leads to the same expression. For timelike currents
the retarded (advanced) field implies that there is no incoming (outgoing)
radiation. This means that the field falls off faster than $1/r$
at null-infinity. Because in our case the charge comes from past and
leaves for future null-infinity, this condition cannot be satisfied.
Moreover, as pointed out, $F_{ab}$ as given by (\ref{eq:ASfield}),
can be obtained from an ultrarelativistic boost of the Coulomb field.
Since the Coulomb field contains no radiation, we claim that (\ref{eq:ASfield})
carries only radiation that is necessarily associated to the light-like
current (\ref{eq:ASCURR}). 

As usual the electric and magnetic parts of $F_{ab}$ may be obtained
by projecting the field-strength and its dual onto a timelike direction
$u^{a}$\[
E=(\partial_{t}\lrcorner F)=-\frac{\sqrt{2}e}{\rho}\delta(u)d\rho\qquad B=(\partial_{t}\lrcorner*F)=-\sqrt{2}e\delta(u)d\phi.\]
(Due to the null-character of the system this decomposition is the
same for any observer with the same transversal $\tilde{x}$-space)
Note that the field has a purely radiative character, i.e. both invariants
of $F_{ab}$ vanish, except at the location of the charge.

\subsection*{2) The Hook-current}

We proceed with the so-called {}``hook-current'', i.e. the electromagnetic
field generated by a point-charge that moves ultrarelativistically
(i.e. the speed of light) and undergoes a sudden change of direction.
This current can also be considered to be the limit of the current
due to a constantly accelerated point-charge in the infinite acceleration
limit. The expression for the hook-current is given by\[
j=j^{n}n+j^{l}l=e\delta^{(2)}(x)(\theta(-u)\delta(v)\partial_{u}+\theta(v)\delta(u)\partial_{v})\]
and consists of two parts: $j^{n}$ resembles a charge $e$ moving
in the direction $n^{a}$ along $v=0=\tilde{x}^{i}$ up to the point
$u=v=0=\tilde{x}^{i}$ where its motion is reversed. $j^{l}$ is the
current in the direction $l^{a}$ along $u=0=\tilde{x}^{i}$. \'{(}See
Fig.1)\\

\begin{picture}(0,0)%
\includegraphics{shock.pstex}%
\end{picture}%
\setlength{\unitlength}{3947sp}%
\begingroup\makeatletter\ifx\SetFigFont\undefined%
\gdef\SetFigFont#1#2#3#4#5{%
  \reset@font\fontsize{#1}{#2pt}%
  \fontfamily{#3}\fontseries{#4}\fontshape{#5}%
  \selectfont}%
\fi\endgroup%
\begin{picture}(2937,3266)(1189,-2735)
\put(4017,-2672){\makebox(0,0)[lb]{\smash{\SetFigFont{14}{16.8}{\rmdefault}{\mddefault}{\updefault}{\color[rgb]{0,0,0}$j^n$}%
}}}
\put(4061,363){\makebox(0,0)[lb]{\smash{\SetFigFont{14}{16.8}{\rmdefault}{\mddefault}{\updefault}{\color[rgb]{0,0,0}$j^l$}%
}}}
\put(4126,-736){\makebox(0,0)[lb]{\smash{\SetFigFont{14}{16.8}{\rmdefault}{\mddefault}{\updefault}{\color[rgb]{0,0,0}$u=0,v>0$}%
}}}
\put(4051,-1636){\makebox(0,0)[lb]{\smash{\SetFigFont{14}{16.8}{\rmdefault}{\mddefault}{\updefault}{\color[rgb]{0,0,0}$v=0,u<0$}%
}}}
\put(3706,-1381){\makebox(0,0)[lb]{\smash{\SetFigFont{12}{14.4}{\rmdefault}{\mddefault}{\updefault}{\color[rgb]{0,0,0}$\tilde{x}$}%
}}}
\end{picture}
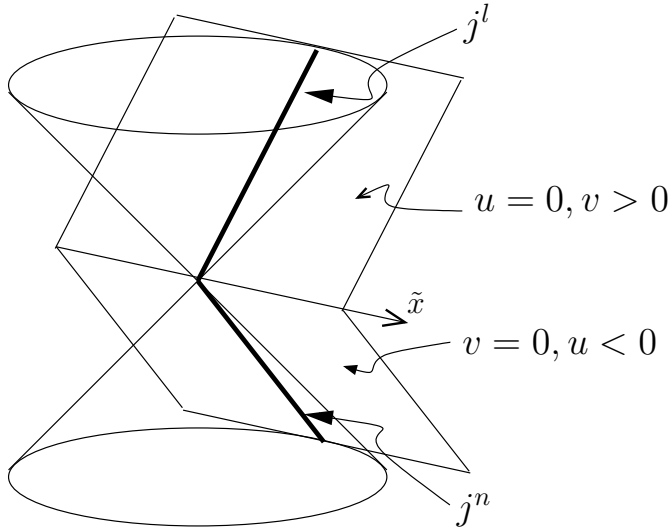
\newline\centerline{Fig.1 {\small The spacetime diagram for the "hook-current"}}
\\
(Charge conservation for the hook-current, i.e. $\partial_{a}j^{a}=0$,
can be easily verified) 

As in the previous case we solve Maxwell's equations in the Lorenz
gauge and adapt the coordinates to the timelike 2-plane spanned by
$l^{a}$ and $n^{a}$. Decomposing the potential\[
A=A^{l}l+A^{n}n\]
 this boils down to two scalar wave-equations\begin{equation}
\partial^{2}A^{l}=4\pi e\delta^{(2)}(x)\delta(u)\theta(v)\qquad\partial^{2}A^{n}=4\pi e\delta^{(2)}(x)\delta(v)\theta(-u).\label{eq:compeq}\end{equation}
The solution of these equations is found by a slight detour: Differentiating
the $A^{l}$-equation with respect to $v$ \[
\partial^{2}\partial_{v}A^{l}=4\pi e\delta^{(4)}(x)\]
and using the identity $\partial^{2}\delta_{+}(x^{2})=-2\pi\delta^{(4)}(x)$
(cf. appendix) immediately gives\begin{equation}
\partial_{v}A^{l}=-2e\delta_{+}(x^{2})\label{eq:inhomeq}\end{equation}
where $\delta_{+}(x^{2})$ denotes the delta-function concentrated
on the future light-cone of the origin in Minkowski space. More rigorously,
it is defined as the functional \begin{equation}
(\delta_{+}(x^{2}),\varphi):=\frac{1}{2}\int_{0}^{\infty}du\int_{0}^{\infty}dv\int d\phi\varphi(u,v,\sqrt{2uv}e_{\rho}),\quad e_{\rho}=(\cos\phi,\sin\phi)\end{equation}
 which arises from {}``splitting'' the action of $\delta(x^{2})$
into its future and past-part. From the above definition we see that
we may symbolically write \[
\delta_{+}(x^{2})=\theta(u)\theta(v)\delta(2uv-\rho^{2}).\]
 Using this identity we may readily integrate $\partial_{v}A^{l}=-2e\delta_{+}(x^{2})$
obtaining \begin{equation}
A^{l}=-e\theta(u)\theta(v)\frac{1}{u}\theta(2uv-\rho^{2}).\label{eq:inhomsol}\end{equation}
 At first glance, due to the appearance of the $\theta(u)/u$ factor,
which has a non-integrable singularity at $u=0$, the result does
not seem to be well defined on all test functions and some kind of
extension has to be used. In the appendix however we show that the
r.h.s. of (\ref{eq:inhomsol}) is a well-defined distribution. Moreover
we proof that $A^{l}$ does indeed satisfy (\ref{eq:inhomeq}).

To find $A^{n}$ we decompose the current on the r.h.s. of (\ref{eq:compeq})
by making use of $\theta(-u)=1-\theta(u):$\[
\partial^{2}A^{n}=4\pi e\delta(v)\delta^{(2)}(\tilde{x})-4\pi e\theta(u)\delta(v)\delta^{(2)}(\tilde{x})\]
Now upon exchanging $u$ and $v$ the second term is precisely of
the form as those in the $A^{l}$-equation. Thus we may follow the
same steps and obtain immediately \[
e\theta(v)\theta(u)\frac{1}{v}\theta(2uv-\rho^{2}).\]
 On the other hand the first term on the r.h.s. of the $A^{n}$-equation
corresponds to a point charge moving at the velocity of light along
$v=0=\tilde{x}^{i}$, i.e. in the direction of $n^{a}$. In the previous
paragraph we have derived the potential for a charge moving in the
direction of $l^{a}$. Making the corresponding changes and adding
both terms we finally obtain for $A^{n}$\[
A^{n}=2e\delta(v)\log\rho+e\theta(v)\theta(u)\frac{1}{v}\theta(2uv-\rho^{2})\]
and total potential for the hook current can be written as\[
A=e\theta(u)\theta(v)\theta(2uv-\rho^{2})(\frac{1}{u}du-\frac{1}{v}dv)-2e\delta(v)\log\rho dv.\]
From this expression it is straightforward to calculate the field-strength
$F_{ab}$, which becomes\[
F=-4e\delta_{+}(x^{2})du\wedge dv-2e\delta_{+}(x^{2})\rho d\rho\wedge(\frac{1}{u}du-\frac{1}{v}dv)-2e\delta(v)\frac{1}{\rho}d\rho\wedge dv.\]

\begin{picture}(0,0)%
\includegraphics{hook.pstex}%
\end{picture}%
\setlength{\unitlength}{3947sp}%
\begingroup\makeatletter\ifx\SetFigFont\undefined%
\gdef\SetFigFont#1#2#3#4#5{%
  \reset@font\fontsize{#1}{#2pt}%
  \fontfamily{#3}\fontseries{#4}\fontshape{#5}%
  \selectfont}%
\fi\endgroup%
\begin{picture}(2279,2559)(289,-1880)
\put(2033,-1448){\makebox(0,0)[lb]{\smash{\SetFigFont{12}{14.4}{\rmdefault}{\mddefault}{\updefault}{\color[rgb]{0,0,0}e}%
}}}
\put(2288,-271){\makebox(0,0)[lb]{\smash{\SetFigFont{12}{14.4}{\rmdefault}{\mddefault}{\updefault}{\color[rgb]{0,0,0}e}%
}}}
\end{picture}
\hspace{0.5cm}\begin{minipage}[b]{0.55\textwidth}%
The physical interpretation is as follows: The point-charge comes
from past null-infinity. At the instant its motion is reversed, a
spherical pulse of radiation is released. This field is concentrated
on the forward light-cone and has the properties of a pure radiation
field, i.e. both invariants of $F_{ab}$ vanish except on the trajectory
of the outgoing charge, where the field diverges. In addition there
is the radiation field concentrated on the null hyperplane of the
incoming charge. (See Fig.2)\end{minipage}%
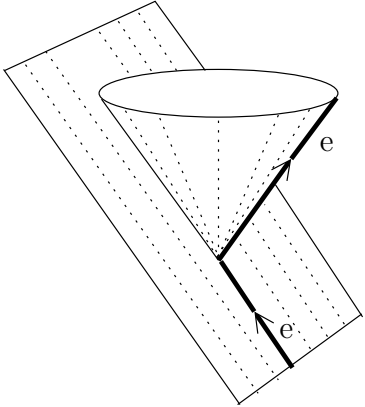
\newline Fig.2 {\small The radiation produced \newline\hspace*{1cm} by the "hook-current"}

\subsection*{3) Head on collision}

From the field of the hook-current we proceed to construct the field
generated by a head-on collision of two ultrarelativistic particles
with opposite charge. The field of the hook-current can be thought
of as the superposition of the field due to the incoming uniformly
moving charge +e and the field that is produced by a pair of charges
+e and -e . These charges emerge from the vertex of the light-cone
and travel with the speed of light together with the spherical pulse,
in opposite directions. By superposing the two configurations on the
forward light-cone the current of the charge -e cancels exactly the
one from +e (which runs in the same direction).

\hspace*{-0.5cm}\begin{minipage}[c]{0.26\columnwidth}%
\begin{picture}(0,0)%
\includegraphics{hcurr.pstex}%
\end{picture}%
\setlength{\unitlength}{3947sp}%
\begingroup\makeatletter\ifx\SetFigFont\undefined%
\gdef\SetFigFont#1#2#3#4#5{%
  \reset@font\fontsize{#1}{#2pt}%
  \fontfamily{#3}\fontseries{#4}\fontshape{#5}%
  \selectfont}%
\fi\endgroup%
\begin{picture}(1725,2041)(391,-1362)
\put(2056,494){\makebox(0,0)[lb]{\smash{\SetFigFont{12}{14.4}{\rmdefault}{\mddefault}{\updefault}{\color[rgb]{0,0,0}$e^+$}%
}}}
\put(2116,172){\makebox(0,0)[lb]{\smash{\SetFigFont{12}{14.4}{\rmdefault}{\mddefault}{\updefault}{\color[rgb]{0,0,0}$e^-$}%
}}}
\put(548,547){\makebox(0,0)[lb]{\smash{\SetFigFont{12}{14.4}{\rmdefault}{\mddefault}{\updefault}{\color[rgb]{0,0,0}$e^-$}%
}}}
\put(391, 89){\makebox(0,0)[lb]{\smash{\SetFigFont{12}{14.4}{\rmdefault}{\mddefault}{\updefault}{\color[rgb]{0,0,0}$e^+$}%
}}}
\put(1996,-1313){\makebox(0,0)[lb]{\smash{\SetFigFont{12}{14.4}{\rmdefault}{\mddefault}{\updefault}{\color[rgb]{0,0,0}$e^+$}%
}}}
\put(473,-1298){\makebox(0,0)[lb]{\smash{\SetFigFont{12}{14.4}{\rmdefault}{\mddefault}{\updefault}{\color[rgb]{0,0,0}$e^-$}%
}}}
\end{picture}
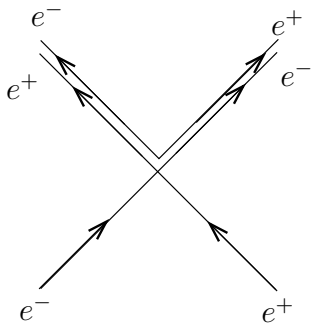
\newline Fig.3 {\small currents of the \\ \hspace*{0.6cm} head-on collision}\end{minipage}%
\hspace{1cm}\begin{minipage}[c]{0.68\columnwidth}%
Now the field for a head-on collision can be constructed as follows:
Consider now the field of two charges +e and -e along the directions
$n^{a}$ and $l^{a}$ respectively\begin{equation}
F_{1}=2e\delta(u)\frac{1}{\rho}d\rho\wedge du-2e\delta(v)\frac{1}{\rho}d\rho\wedge dv.\label{eq:2AS}\end{equation}
Superimpose to the field associated with these charges the field of
a pair of charges $-e$, $+e$ starting from the point of collision
of the incoming charges (as considered above) \begin{equation}
F_{2}=-2e\delta_{+}(x^{2})\left[2du\wedge dv+\rho d\rho\wedge(\frac{du}{u}-\frac{dv}{v})\right]\label{eq:ForwCone}\end{equation}
\end{minipage}%
\newline\noindent Thus the total field is the sum of the fields $F_{1}$
and $F_{2}$. On the forward light-cone with vertex at the collision
event, the currents cancel and a pure radiation field with no sources
remains. 

\vspace{0.5cm}
\begin{picture}(0,0)%
\includegraphics{headon.pstex}%
\end{picture}%
\setlength{\unitlength}{3947sp}%
\begingroup\makeatletter\ifx\SetFigFont\undefined%
\gdef\SetFigFont#1#2#3#4#5{%
  \reset@font\fontsize{#1}{#2pt}%
  \fontfamily{#3}\fontseries{#4}\fontshape{#5}%
  \selectfont}%
\fi\endgroup%
\begin{picture}(4737,3894)(214,-3455)
\put(4951,-1846){\makebox(0,0)[lb]{\smash{\SetFigFont{12}{14.4}{\rmdefault}{\mddefault}{\updefault}{\color[rgb]{0,0,0}support of $F_1$}%
}}}
\put(4666,-301){\makebox(0,0)[lb]{\smash{\SetFigFont{12}{14.4}{\rmdefault}{\mddefault}{\updefault}{\color[rgb]{0,0,0}support of $F_2$}%
}}}
\end{picture}
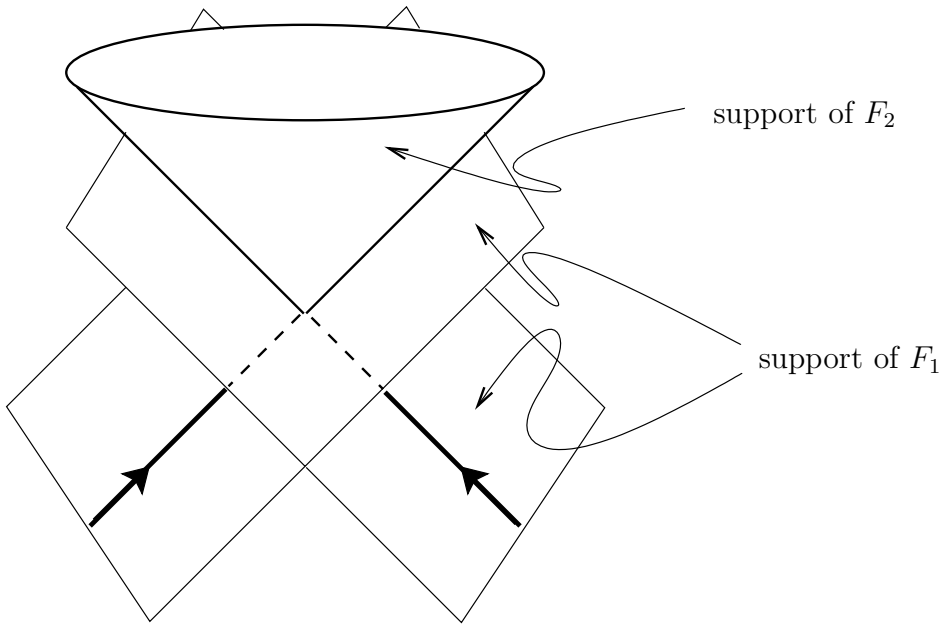
\newline Fig.4 {\small The EM field of two colliding charges. The field consists \newline\hspace*{1cm}of the incoming part $F_1$ plus the radiation $F_2$ produced \newline\hspace*{1cm}$F_1$by the impact}\vspace{0.5cm}\newline\noindent
In what follows we discuss the radiation (\ref{eq:ForwCone}) produced
by the head-on collision in more detail. Since \[
F_{2}\propto\delta_{+}(x^{2})=\frac{1}{2r}\delta(t-r),\,\,\, r^{2}=\rho^{2}+z^{2}\]
this part is an outgoing spherical pulse centered around the collision
point. Its overall strength decreases as $1/r$, but the field is
not spherically symmetric. However, since the physical situation is
axisymmetric about the spatial direction of the current ($z$-axis)
the field carries this symmetry. Decomposing $F_{2}$ into its electric
and magnetic parts with respect to $\partial_{t}$ we find\[
\partial_{t}\lrcorner F_{2}=-\frac{4e}{\rho}\delta_{+}(x^{2})(\rho dz-zd\rho)\]
i.e.\[
E=-\frac{4e}{\rho}\delta_{+}(x^{2})(\rho e_{z}-ze_{\rho})\]
and from \[
*F_{2}=-2e\delta_{+}(x^{2})\left[-2d^{2}\tilde{x}+(\frac{du}{u}+\frac{dv}{v})\rho^{2}d\phi\right]\]
and\[
\partial_{t}\lrcorner*F_{2}=-2e\delta_{+}(x^{2})\left[\frac{1}{\sqrt{2}}(\frac{1}{u}+\frac{1}{v})\rho^{2}d\phi\right]=-4e\delta_{+}(x^{2})td\phi\]
i.e.

\[
B=-4e\delta_{+}(x^{2})\frac{\sqrt{\rho^{2}+z^{2}}}{\rho}e_{\phi}\]
where $e_{\rho},e_{\phi},e_{z}$ denote the orthonormal vectors along
the corresponding coordinates. In order to show that both Maxwell
invariants vanish we {}``smear'' $E$ and $B$ with the same time-dependent
test-function $\varphi(t)$, i.e.\begin{eqnarray}
E_{\varphi} & := & (E,\varphi(t))=-\frac{4e}{\rho}(\delta_{+},\varphi)(\rho e_{z}-ze_{\rho})\nonumber \\
B_{\varphi} & := & (B,\varphi(t))=-\frac{4e}{\rho}(\delta_{+},\varphi)\sqrt{\rho^{2}+z^{2}}e_{\phi}\label{eq:smeared-fields}\end{eqnarray}
which are both regular functionals of the spatial variables. It is
now easy to see that\[
E_{\varphi}^{2}=16e^{2}(\delta_{+},\varphi)^{2}\frac{\rho^{2}+z^{2}}{\rho^{2}}=B_{\varphi}^{2}\qquad E_{\varphi}\cdot B_{\varphi}=0\]
for arbitrary smearing functions $\varphi$. This holds everywhere
except for $\rho=0$ where there is an additional $1/\rho$ singularity.
However, it is precisely along the two lines on the light cone where
$F_{1}$ contributes. Adding $F_{1}$ shows that the $1/\rho$ singularity
cancels and the total field is of the form\begin{equation}
F\propto\delta_{+}(x^{2})du\wedge dv;\qquad\mbox{close\, to\, }\rho=0,x^{2}=0,t>0.\label{eq:roughfield}\end{equation}
This claim can be strengthened in a rigorous manner using smeared
field strengths, i.e.\begin{eqnarray*}
F_{1}(\varphi) & := & (F_{1},\varphi(t))=2\sqrt{2}e\left(\frac{1}{\rho}d\rho\wedge du\varphi(r\cos\theta)-\frac{1}{\rho}d\rho\wedge dv\varphi(-r\cos\theta)\right)\\
F_{2}(\varphi) & := & (F_{2},\varphi(t))=-\frac{2e}{r}\varphi(r)\left(du\wedge dv+\right.\\
 &  & \left.\frac{1}{\sqrt{2}\sin\theta}\left[(1+\cos\theta)d\rho\wedge du-(1-\cos\theta)d\rho\wedge dv\right]\right)\\
 &  & \qquad\qquad\qquad supp(\varphi)=\left\{ t|t>t_{0}\geq0\right\} .\end{eqnarray*}
Adding $F_{1}(\varphi)$ and $F_{2}(\varphi)$ toghether and taking
the limits $\theta\to0$,$\pi$ one immedeately finds \begin{eqnarray*}
F_{1}(\varphi) & \overset{\theta\to0}{\sim} & \frac{2\sqrt{2}e}{r\sin\theta}\varphi(r)d\rho\wedge du,\\
F_{2}(\varphi) & \overset{\theta\to0}{\sim} & -\frac{2e}{r}\varphi(r)du\wedge dv-\frac{2\sqrt{2}e}{r\sin\theta}\varphi(r)d\rho\wedge du\\
F_{1}(\varphi) & \overset{\theta\to\pi}{\sim} & -\frac{2\sqrt{2}e}{r\sin\theta}\varphi(r)d\rho\wedge dv,\\
F_{2}(\varphi) & \overset{\theta\to\pi}{\sim} & -\frac{2e}{r}\varphi(r)du\wedge dv+\frac{2\sqrt{2}e}{r\sin\theta}\varphi(r)d\rho\wedge dv\end{eqnarray*}
which proves the qualitative claim (\ref{eq:roughfield}). 

From the construction it is evident that the total field satisfies
the source-free Maxwell equations everywhere to the future of the
collision.

\subsection*{Conclusion}

In this work we have explicitly constructed the field due to a charge
moving at ultrarelativistic velocities, which undergoes a sudden change
of direction, the so-called hook-current. Upon time-reflection of
the outgoing charge one obtains the field of the head-on collision,
i.e. the classical analogue of pair annihilation leaving a pure radiation
field behind.

From a distributional point of view and due to the null-character
of the situation our construction is rather delicate, which manifests
itself in the fact that the formal expression for the potential seems
to contain a non-integrable singularity thus not giving rise to functional
defined on all of test-function space. However a careful analysis
shows that it is nevertheless possible to construct a well-defined
distribution from the formal expression which satisfies the required
differential equation. Recently the problem of colliding ultrelativistic
massless particles and the possible formation of black-holes has attracted
some interest. Podolsky et al. \cite{PodGr} proposed the null limit
of the C-metric \cite{KinWal} as scenario describing the headon-collision
of two ultrarelativistic black holes. However, because of the existence
of singular strut between the black holes we believe that this is
not the appropriate physical situation. The gravitational analogue
of the head-on collision is more realistically considered by Giddings
et al.\cite{Gid}. We do hope that our work may shed some light on
the corresponding distributional techniques required in the gravitational
context.

\newpage

\subsection*{Appendix}

This appendix is devoted to the distributional aspects of our work,
namely the definition of $\delta_{+}(x^{2})$ and the construction
of the solution of $\partial_{u}f=\delta_{+}(x^{2})$. \\
Using the definition of $\delta_{+}$\begin{equation}
(\delta_{+}(x^{2}),\varphi):=\frac{1}{2}\int_{0}^{\infty}du\int_{0}^{\infty}dv\int d\phi\varphi(u,v,\sqrt{2uv}e_{\rho}),\end{equation}
 and its symbolic form \[
\delta_{+}(x^{2})=\theta(u)\theta(v)\delta(2uv-\rho^{2}),\]
 allows a (formal) integration of $\partial_{u}f=\delta_{+}(x^{2})$
\begin{equation}
f=\theta(u)\theta(v)\frac{1}{v}\theta(2uv-\rho^{2}).\end{equation}
 At first glance, due to the appearance of the $\theta(v)/v$ factor,
which has a non-integrable singularity at $v=0$, the result does
not seem to be well defined on all test functions and some kind of
extension has to be used. Let us briefly recall how this is done for
$\theta(v)/v$. Obviously \[
(\frac{\theta(v)}{v},\varphi)=\int_{0}^{\infty}dv\frac{1}{v}\varphi(v)\]
 is only defined on test functions that vanish at $v=0$. An extension
to all of test function space is achieved by mapping $\varphi$ onto
an integrable function with this property, i.e. \[
([\frac{\theta(v)}{v}],\varphi):=\int_{0}^{\infty}dv\frac{1}{v}(\varphi(v)-\theta(1-v)\varphi(0)),\]
 which coincides with the naive definition when $\varphi(0)=0$. The
subtraction term depends explicitly on the neighborhood of zero (which
we have chosen to have radius one). Let us now turn back to $\theta(u)(\theta(v)/v)\theta(uv-\rho^{2})$,
i.e. \begin{eqnarray*}
(\theta(u)\theta(v)\frac{1}{v}\theta(2uv-\rho^{2}),\varphi) & = & \int_{0}^{\infty}du\int_{0}^{\infty}dv\frac{1}{v}\int_{0}^{\sqrt{2uv}}\rho d\rho d\phi\varphi(u,v,\rho e_{\rho})\\
 & = & \int_{0}^{\infty}du\int_{0}^{\infty}dv\frac{1}{v}\Phi(u,v)\\
\mbox{with} &  & \Phi(u,v):=\int_{0}^{\sqrt{2uv}}\rho d\rho d\phi\varphi(u,v,\rho e_{\rho})\end{eqnarray*}
 Our previous considerations suggest to take a closer look at the
small-$v$ behavior of $\Phi$\begin{eqnarray*}
\Phi(u,v) & = & \Phi(u,0)+v\partial_{v}\Phi(u,0)+\mathcal{O}(v^{2})\\
 & = & v(2\pi u\varphi(u,0,0,0))+\mathcal{O}(v^{2}),\end{eqnarray*}
 which tells us that $\Phi$ has a zero at $v=0$ and therefore in
the light of the extension-argument \[
(\theta(u)\theta(v)\frac{1}{v}\theta(2uv-\rho^{2}),\varphi):=\int_{0}^{\infty}du\int_{0}^{\infty}dv\frac{1}{v}\int_{0}^{\sqrt{2uv}}\rho d\rho d\phi\varphi(u,v,\rho e_{\rho})\]
 is well defined on all of test function space. Stated differently
$\theta(u)(\theta(v)/v)\theta(uv-\rho^{2})$ is a well-defined distribution.

Let us now consider the action of the d'Alembertian $\partial^{2}$
on $(\theta(u)/u)\theta(v)\theta(2uv-\rho^{2})$. As a warm-up and
to get acquainted with the neccessary techniques let us first show
that $\theta(u)\theta(v)\delta(2uv-\rho^{2})$ is the retarded Green-function
of $\partial^{2}$, i.e.\begin{eqnarray*}
(\partial^{2}(\theta(u)\theta(v)\delta(2uv-\rho^{2})),\varphi) & = & (\theta(u)\theta(v)\delta(2uv-\rho^{2}),\partial^{2}\varphi)\\
 & = & \frac{1}{2}\int_{0}^{\infty}du\int_{0}^{\infty}dv\int_{0}^{2\pi}d\phi\,\partial^{2}\varphi(u,v,\sqrt{2uv}e_{\rho}).\end{eqnarray*}
Taking into account the coordinate decomposition of $\partial^{2}=-2\partial_{u}\partial_{v}+\delta^{ij}\partial_{i}\partial_{j}$
we split the last expression into two parts\begin{eqnarray*}
 &  & u,v-part=\frac{1}{2}\int_{0}^{\infty}du\int_{0}^{\infty}dv\int_{0}^{2\pi}d\phi\,\partial_{u}\partial_{v}\varphi(u,v,\sqrt{2uv}e_{\rho})\\
 &  & \qquad=\frac{1}{2}\int_{0}^{\infty}du\int_{0}^{\infty}dv\left\{ \partial_{u}\int_{0}^{\infty}d\phi\partial_{v}\varphi-\sqrt{\frac{v}{2u}}\int_{0}^{2\pi}d\phi e_{\rho}^{i}\partial_{v}\partial_{i}\varphi\right\} \\
 &  & \qquad=\pi\varphi(0)-\frac{1}{2}\int_{0}^{\infty}du\int_{0}^{\infty}dv\sqrt{\frac{v}{2u}}\left\{ \partial_{v}\int_{0}^{2\pi}d\phi\, e_{\rho}^{i}\partial_{i}\varphi-\sqrt{\frac{u}{2v}}\int_{0}^{2\pi}d\phi e_{\rho}^{i}e_{\rho}^{j}\partial_{i}\partial_{j}\varphi\right\} \\
 &  & \qquad=\pi\varphi(0)+\frac{1}{4}\int_{0}^{\infty}du\int_{0}^{\infty}dv\left\{ \frac{1}{\sqrt{2uv}}\int_{0}^{2\pi}d\phi\, e_{\rho}^{i}\partial_{i}\varphi+\int_{0}^{2\pi}d\phi e_{\rho}^{i}e_{\rho}^{j}\partial_{i}\partial_{j}\varphi\right\} \end{eqnarray*}

\begin{eqnarray*}
 &  & i,j-part=\frac{1}{2}\int_{0}^{\infty}du\int_{0}^{\infty}dv\int_{0}^{2\pi}d\phi\,\delta^{ij}\partial_{i}\partial_{j}\varphi(u,v,\sqrt{2uv}e_{\rho})\\
 &  & \qquad=\frac{1}{2}\int_{0}^{\infty}du\int_{0}^{\infty}dv\int_{0}^{2\pi}d\phi\,(e_{\rho}^{i}e_{\rho}^{j}+e_{\phi}^{i}e_{\phi}^{j})\partial_{i}\partial_{j}\varphi\\
 &  & \qquad=\frac{1}{2}\int_{0}^{\infty}du\int_{0}^{\infty}dv\left\{ \int_{0}^{2\pi}d\phi e_{\rho}^{i}e_{\rho}^{j}\partial_{i}\partial_{j}\varphi+\frac{1}{\sqrt{2uv}}\int_{0}^{2\pi}d\phi e_{\phi}^{i}\partial_{\phi}\partial_{i}\varphi\right\} \\
 &  & \qquad=\frac{1}{2}\int_{0}^{\infty}du\int_{0}^{\infty}dv\left\{ \int_{0}^{2\pi}d\phi e_{\rho}^{i}e_{\rho}^{j}\partial_{i}\partial_{j}\varphi+\frac{1}{\sqrt{2uv}}\int_{0}^{2\pi}d\phi e_{\rho}^{i}\partial_{i}\varphi\right\} \end{eqnarray*}
Adding the u,v--part (multiplied with a factor $-2$) to the i,j--part
finally gives\[
(\partial^{2}(\theta(u)\theta(v)\delta(2uv-\rho^{2})),\varphi)=-2\pi\varphi(0)=-2\pi(\delta^{(4)}(x),\varphi)\]
thus proving the desired result. Proceeding in an analoguous manner
we will evaluate the d'Alembertian on $(\theta(u)/u)\theta(v)\theta(2uv-\rho^{2})$,
i.e.\begin{eqnarray*}
(\partial^{2}(\theta(u)\theta(v)\frac{1}{u}\theta(2uv-\rho^{2})),\varphi) & = & (\theta(u)\theta(v)\frac{1}{u}\theta(2uv-\rho^{2}),\partial^{2}\varphi)\\
 & = & \int_{0}^{\infty}du\int_{0}^{\infty}dv\frac{1}{u}\int_{0}^{\sqrt{2uv}}\rho d\rho\int_{0}^{2\pi}d\phi\,\partial^{2}\varphi(u,v,\rho e_{\rho})\end{eqnarray*}
Splitting into $u,v$- and $i,j$--part yields\begin{eqnarray*}
 &  & u,v-part=\int_{0}^{\infty}du\int_{0}^{\infty}dv\frac{1}{u}\int_{0}^{\sqrt{2uv}}\rho d\rho\int_{0}^{2\pi}d\phi\,\partial_{u}\partial_{v}\varphi(u,v,\rho e_{\rho})\\
 &  & \qquad=\int_{0}^{\infty}du\int_{0}^{\infty}dv\left\{ \frac{1}{u}\partial_{v}\int_{0}^{\sqrt{2uv}}\rho d\rho\int_{0}^{2\pi}d\phi\,\partial_{u}\varphi-\int_{0}^{2\pi}d\phi\,\partial_{u}\varphi\right\} \\
 &  & \qquad=-\int_{0}^{\infty}du\int_{0}^{\infty}dv\int_{0}^{2\pi}d\phi\,\partial_{u}\varphi\\
 &  & \qquad=2\pi\int_{0}^{\infty}dv\varphi(0,v,0)+\int_{0}^{\infty}du\int_{0}^{\infty}dv\frac{2v}{2\sqrt{2uv}}\int_{0}^{2\pi}d\phi\, e_{\rho}^{i}\partial_{i}\varphi\end{eqnarray*}
 and \begin{eqnarray*}
 &  & i,j-part=\int_{0}^{\infty}du\int_{0}^{\infty}dv\frac{1}{u}\int_{0}^{\sqrt{2uv}}\rho d\rho\int_{0}^{2\pi}d\phi\,\partial^{i}\partial_{i}\varphi(u,v,\rho e_{\rho})\\
 &  & \qquad=\int_{0}^{\infty}du\int_{0}^{\infty}dv\frac{1}{u}\int_{0}^{2\pi}d\phi\,\sqrt{2uv}e_{\rho}^{i}\partial_{i}\varphi\end{eqnarray*}
Adding u,v--part (multiplies by a factor $-2$) and $i,j$--part finally
gives\[
(\partial^{2}(\theta(u)\theta(v)\frac{1}{u}\theta(2uv-\rho^{2})),\varphi)=-4\pi\int_{0}^{\infty}dv\,\varphi(0,v,0)=-4\pi(\theta(v)\delta(u)\delta^{(2)}(x),\varphi),\]
which once again is the desired result.

\end{document}